\documentclass{amsart}
\usepackage{amsmath}%
\usepackage{amsfonts}%
\usepackage{amssymb}%
\usepackage{graphicx}
\theoremstyle{plain}

\numberwithin{equation}{section}

\def\beq{\begin{eqnarray}}
\def\eeq{\end{eqnarray}}

\def\H{\mathcal{H}}

\def\S{\mathcal{S}}
\def\M{\mathcal{M}}

\def\rr{{\mathbb{R}}}

\begin{document}
\title{On the Role of Quantum Events in Double-Slit Experiments}
\author{R. Schuster}
\address[Auckland, New Zealand]
{Auckland, New Zealand}%
\email{rschuster.physics@gmail.com}%
\date{April 10, 2011}

\begin{abstract}
We formulate the Schr\"odinger equation as the equation of motion of a small  external influence
which serves as the initial boundary condition of a physical system in classical laboratory space. The Hilbert space of possible external influences for a given
physical system is then equivalent to the Hilbert space of a quantum system without spin. 
We discuss the double-slit experiment in the context of this approach and show that wave-particle dualism 
reduces to the choice of a basis
in the geometrical construction of the Hilbert space of small external influences. 
\end{abstract}
\maketitle

\section{Introduction}
The current development in mastering the creation of smallest objects rises a new era of tests, which challenges quantum theory on it's most fundamental level \cite{Aspelmeyer1}. The nature of these tests will move hypothetical Gedankenexperiments into the reach of experimental confirmation,
and
shine new light on the interpretation 
issues, which plague quantum mechanics from it's very beginnings.
As this recent progress tackles the issues at an interpretation level only, it
needs to be accompanied by a structural rethinking of the theoretical fundamental concepts to gain substantially new insight. 

We approach quantum mechanics by 
using the mathematical formalism to variational problems of Gelfand and Fomin \cite{Gelfand1}, and
investigate the time evolution of an external influence,
which serves as the initial boundary condition 
of a classical physical system given 
with free initial point.

The necessary and sufficient condition for the external influence to 
serve as a boundary condition for the classical system in the entire time interval
is given by the Hamilton-Jacobi equation.
We use the algebraic one-to-one correspondence 
between the elements of the algebra of transformations and the elements of the corresponding covering group 
near the identity to
derive the  Schr\"odinger equation  as the equation of motion of a field theory
in classical configuration space
for a  small external influence.
The 
Hilbert space of quantum states in energy eigenstate representation corresponds, then, to
the set of possible solutions of the above field theory, where 
 only the
 set
of basis vectors of this solution space is observable in classical space
due to the role of the Hamilton-Jacobi equation as the necessary and sufficient condition.

The distinguished role of the energy eigenvalue representation enables us to apply our formalism to the experimental results 
of a double-slit experiment without modification and explains the continuity of the interference pattern
in a natural way. This is in contrast to  standard quantum mechanics, where one needs
 to extend the formalism by alternative methods, e.{}g.{} the positive operator valued measure (POVM) approach \cite{deMuynck1}, \cite{Peres1}, to formulate the observed results of the double-slit experiment.
Our results allow us to show in particular that the wave-particle dualism observed in the double-slit experiment reduces to the experimental realization of the choice of a basis in a degenerate Hilbert space
 constituted by the energy eigenfunctions emanating from each slit.
 
Another particular aspect in our approach is the close connection between the quantum object - the small external influence
as a boundary condition - and the classical laboratory configuration - the influenced
classical physical system, which emphasizes the geometry of the classical configuration space at the time
of the appearance of the initial influence in the classical system.  Our view shifts the quantum property, particle or wave,
of a quantum object to the property of a quantum event, the event of appearance in classical laboratory space with configuration space geometry given at this point in time. This explains the observed behaviour of delayed choice experiments if we
 regard the contact of the quantum object with the screen/counter as the  quantum event of significance.

 From a formal perspective our derivation is closely related to Schr\"odinger's 
 first note \cite{Schroedinger1}. But in contrast to Schr\"odinger,
who abandoned this approach by lack of physical justification, we have an intuitive physical interpretation and mathematically rigorous derivation which gives new insight into the 
nature of quantum mechanics.

\section{Quantum System as External Influence}

Let us consider a physical system with 
action 
\beq
\\
 \int_a^b L(t, q_1, \ldots, q_n,  \dot{ q}_1, \ldots, \dot{q}_n ) dt + 
g_1(a, q_1(a), \ldots, q_n(a)) + g_2(b, q_1(b), \ldots, q_n(b)).
\nonumber
\label{eq:VarProb_1}
\eeq
and variable endpoints $a$ and $b$%
\footnote
{
Although this action
 is unusual in fundamental physics it is common in applied science in its generalized form
 given in optimal theory, 
see e.{}g.{} \cite{Kirk1}.
The task is to find an admissible control function $u(t)$ 
which causes the system to follow an admissible trajectory $x(t)$ subject to the dynamical equation
$$
\dot{x} = a(x(t), u(t),t) 
$$
and optimizes the action
$$
I(u) = \varphi(x(t_f),t_f,x(t_i),t_i) + \int_{t_i}^{t_f} L (x(t), u(t),t) dt.
$$
with boundary condition $\varphi(x(t_f),t_f,x(t_i),t_i)$.
}.
The action of this system differs from the familiar action
\beq
\\
\;\;\;\;\;\;\;\;\;\; \int_a^b L(t, q_1, \ldots, q_n,  \dot{ q}_1, \ldots, \dot{q}_n ) dt 
\nonumber
\label{eq:VarProb_fam} 
\eeq
by the additional terms $g(a, q_1(a), \ldots, q_n(a))$ and $g_2(b, q_1(b), \ldots, q_n(b))$, 
which are functions of the boundary points only.
As known from elementary theory these functions serve as the boundary conditions which single out a definite solution and 
represent the  external influence on a physical system described
by the Lagrangian (\ref{eq:VarProb_fam}).
As an example, suppose the above Lagrangian (\ref{eq:VarProb_fam}) represents an experimental set up, which is kicked off and changes to its initial state at the free initial time $a$.
Then, this Lagrangian describes the evolution of the classical  physical system over time $t$ in the laboratory. 
For reasons which will be clear later in the text
we will call 
the system described by (\ref{eq:VarProb_fam}) the laboratory system  in the following.

The dynamics of the laboratory system is given by the Euler equations\footnote
{
In the following we follow closely the line of thoughts in \cite{Gelfand1}, Chap.{} 6 and omit proofs which can be found in
this book.
}
\beq
\frac{\partial L}{\partial q_i} - \frac{d}{d t} \left(\frac{\partial L}{\partial \dot{q_i}}\right) = 0,
\label{eq:Euler}
\eeq
and the additional boundary conditions
\beq
\left. \frac{\partial L}{\partial \dot{q_i}}\right|_{t = a} - \frac{\partial g_1(a,q)}{\partial q_i} = 0
\label{eq:InitCond}
\eeq
and 
\beq
\left. \frac{\partial L}{\partial \dot{q_i}}\right|_{t = b} - \frac{\partial g_2(b,q)}{\partial q_i} = 0.
\label{eq:FinalCond}
\eeq
Instead of investigating the equation of motions of the laboratory system let us concentrate on the initial boundary conditions and write the external influence 
\beq
g_1(a,q) = g(t,q)\bigg|_{t = a}
\eeq
as a function of time evaluated at $a$ and omit the subscript in the following. 
Note similar results can be observed for the boundary conditions at the terminal point, which we will ignore in the following by assuming that $g_2(b, q_1(b), \ldots, q_n(b)) = 0$ at the terminal point $b$.

Condition (\ref{eq:InitCond}) relates the external initial influence $g(a,q)$
 to the momentum of the laboratory system
\beq
\left. p_i(t,q,\dot{q})\right.\bigg|_{t = a} = \left. \frac{\partial g(t,q)}{\partial q_i}\right.\bigg|_{t = a}
\label{eq:MomDist}
\eeq
at the initial point $a$, where we
 define the momentum as usual as the velocity derivative of the Lagrangian
\beq
p_i(t,q,\dot{q}) =  \frac{\partial L(t,q,\dot{q})}{\partial \dot{q_i}}.
\label{eq:DefMom}
\eeq
Relations (\ref{eq:MomDist}) and (\ref{eq:DefMom}) allow us to rewrite the boundary conditions as functions of configuration space
variables 
\beq
\dot{q_i}(a) = \varphi_i(q)\bigg|_{t = a} \ \ \ \ \ \ \ (i= 1,\ldots,n),
\label{eq:NatBoundary}
\eeq
which can be thought to assign a direction to every point in the hyperplane $t=a$ and allows us to define a family 
of boundary conditions as follows.
The family of boundary conditions
\beq
\dot{q_i}(t) = \varphi_i(t,q) \ \ \ \ \ \ \ (i= 1,\ldots,n)
\eeq
imposed for every $t \in [a,b]$ represents a field of the functional (\ref{eq:VarProb_fam}) if 
\begin{itemize}
	\item  there exists a function $g(t,q)$ such that 
	\beq
	\left. p_i(t,q,\varphi(q))\right.\bigg|_{t = a} = \left. \frac{\partial g(t,q)}{\partial q_i}\right.\bigg|_{t = a}.
	\label{eq:SelfAdj}
	\eeq
  That is, the external influence represented by $g(t,q)$ is in contact with the laboratory system imposing the initial momentum
  $p_i(a,q(a),\dot{q}(a))$ at time $a$.
	\item every extremal, i.e. solution of the Euler equations, satisfying the boundary conditions
	\beq
	\dot{q_i}(t_1) = \varphi^{(1)}_i(q)\bigg|_{t = t_1}
	\label{eq:Trace1}
	\eeq
	also satisfies the boundary conditions
	\beq
	\dot{q_i}(t_2) = \varphi^{(2)}_i(q)\bigg|_{t = t_2}
	\label{eq:Trace2}
	\eeq
	at the different point in time $t_2$ and vice versa, i.{}e.{} the boundary conditions are traceable in time. Boundary conditions of this
	type are called consistent.
\end{itemize}

Thus, the influence function $g(t,q)$ acts as a kind of potential in configuration space for the family of boundary conditions.
Note the potential  $g(t,q)$ is given at specific points in time where the configuration space points should be regarded as
a set of points at this given time and not parameterized by $t$ as (\ref{eq:SelfAdj}) is a relation for equal times.
Since
boundary conditions describe the external influence of an unknown source to a physical system, which 
 can be completely  unrelated at different points in time, this means physically spoken, that
we restrict our investigation to the class of external influences which are historically traceable  and have a 
configuration space representation
of fields which are derivable from a potential.

 Then, the natural question arises which condition must fulfill the external influence function $g(t,q)$
  to keep the ability  to kick off the laboratory system described by (\ref{eq:VarProb_fam})
 at an arbitrary point  $t \in [a,b]$ in time.
Gelfand and Fomin, \cite{Gelfand1} p.{} 146, showed that the necessary and sufficient condition, called consistency condition,
is the Hamilton-Jacobi equation
\beq
\frac{\partial g(t,q)}{\partial t} + H\left(t,q,
\frac{\partial g(t,q)}{\partial q_1}, \ldots, \frac{\partial g(t,q)}{\partial q_n}
\right) = 0,
\label{eq:HamJac}
\eeq
with Hamilton function\footnote
{
Note the following derivation based on this equation is of a local character as the Jacobian for the transformation to the generalized momentum leading to (\ref{eq:HamJac}) is only locally valid as pointed out in footnote 2 of \cite{Gelfand1}, p68.
That is, we will have an explicitly local derivation of quantum mechanics in contrast to the standard approach to quantum mechanics.
} $H(t,q,p)$.
Thus, the set of external influence functions $\{g_i\}$, which constitute the  solutions of the Hamilton-Jacobi equation (\ref{eq:HamJac}) in local coordinates,  are the generators of the canonical transformation in the laboratory system
and, therefore,  the external influence function $g \in \{g_i\}$ is an element of a Lie algebra.

Suppose we have an external influence $g$ which causes a very small transformation in laboratory space. Then, we can
use the isomorphism between  
elements in the neighbourhood  of 0 of a Lie algebra $\mathfrak g$ and 
the Lie subgroup $G_0$ of elements connected to the identity established via the inverse
exponential map
\beq
g = \ln \psi
\eeq
and
do the ansatz 
\beq
g(t,q) = k \ln \psi(t,q)
\label{eq:SubstLog}
\eeq
in local coordinates.
Physically spoken, this means we want to investigate the object itself represented as an element of the group of objects
and
are not interested in the information space generated by the external influence given by the tangential
space, which is isomorphic to the Lie algebra.
Then 
 (\ref{eq:HamJac}) reads
\beq
\frac{k}{\psi(t,q)} \frac{\partial \psi(t,q)}{\partial t} + H\left(q, \frac{k}{\psi(t,q)} \frac{\partial \psi(t,q)}{\partial q} \right)  = 0,
\label{eq:TimeDep_0}
\eeq
where the constant $k$ is assumed to be small to guarantee the smallness of $g(t,q)$ over  entire
configuration space and time. In fact we know from experiment that nature has chosen a very small value 
\beq
k \propto \hbar
\eeq
with Planck's constant $\hbar$.

For conservative systems $H = E$ relation (\ref{eq:TimeDep_0}) decouples into
\beq
\frac{k}{\psi(t,q)} \frac{\partial \psi(t,q)}{\partial t} + E = 0
\label{eq:TimeDep}
\eeq
and
\beq
- E + H\left(q, \frac{k}{\psi(t,q)} \frac{\partial \psi(t,q)}{\partial q} \right) = 0.
\label{eq:ConsPart}
\eeq
The first equation (\ref{eq:TimeDep}) reads 
\beq
k \frac{\partial \psi}{\partial t} = E \psi,
\eeq
where we suppress the arguments in the following, and represents the time evolution of the external influence, which is in contact with laboratory space with energy $E$.

The second condition (\ref{eq:ConsPart}) can be reformulated in terms of the kinetic and potential energy of the Hamiltonian 
$H = T + V$ in laboratory space
\beq
k^2 \left(\frac{\partial \psi}{\partial q}\right)^2 + V \psi^2 - E \psi^2 = 0.
\label{eq:ConsConClass2}
\eeq

To describe the dynamics of the external influence in laboratory space
we interpret the left side of this equation as the Lagrangian 
\beq
L(\psi,\partial_q \psi) = k^2 \left(\frac{\partial \psi}{\partial q}\right)^2 + V \psi^2 - E \psi^2
\label{eq:ConsConClass3}
\eeq
of a field theory\footnote
{
Remember in the derivation of the consistency condition (\ref{eq:HamJac}), which is our point of departure, the connection between the external influence function and the momentum
in laboratory space is given by the equal time relation 
$$
\left. p_i(t,q,\dot{q})\right.\bigg|_{t = a} = \left. \frac{\partial g(t,q)}{\partial q_i}\right.\bigg|_{t = a}.
$$
That is, the  function $g(t,q)$, and $\psi(t,q)$, is considered for an arbitrary but fixed point in time in laboratory space and 
the  points $q$ should be considered as the set of points constituting the configuration space at this point in time. 
As the set of solutions of the laboratory space
dynamics of the field $\psi(q)$ will result in the Hilbert space of states,
this view is similar to the standard approach in quantum mechanics where the Hilbert space of states is derived for a fixed point in time
and the time-dependent Schr\"odinger equation is postulated additionally to describe the time evolution of the system.
}
 in $\psi$
and its space derivative $\partial_q \psi = {\partial \psi}/{\partial q}$,
which is equivalent to the variational problem 
\beq
L(\psi,\partial_q \psi) = k^2 \left(\frac{\partial \psi}{\partial q}\right)^2 + V \psi^2
\label{eq:ConsConClass4}
\eeq
with constraint
\beq
 \int \left|\psi\right|^2 dq = 1
 \label{eq:Constraint4}
\eeq
where the integration  is over entire configuration space and the energy constant $E$ plays the role of 
the Lagrange multiplier.

The equation of motion can be easily read from (\ref{eq:ConsConClass3}) and results into the stationary 
equation
\beq
H\left( q, \frac{ \hbar}{i} \frac{\partial}{\partial q}\right) \psi = E \psi
\label{eq:SG_FinalForm}
\eeq
with $k = \hbar/i$ and Hamiltonian $H$ in operator notation.
Insertion of the above equation
into (\ref{eq:TimeDep})
leads to the time-dependent Schr\"odinger equation
\beq
i \hbar \frac{\partial \psi}{\partial t} = H \psi
\label{eqTDSG}
\eeq
for the external influence $\psi$ and we can identify the Schr\"odinger equations as the dynamics of a very small external influence
fulfilling the consistency condition (\ref{eq:HamJac}), which is necessary and sufficient to keep the external influence in contact with laboratory space.

Moreover, it is well known, see e.{}g.{} \cite{Courant1}, that the set of solutions of the above variational problem
form an orthonormal basis of a Hilbert space of square integrable functions over laboratory space $\M$, which 
is subject to the experimental setting by virtue of the consistency condition (\ref{eq:HamJac}). 
Thus, the  external influences connected to laboratory space
do not only fulfil the Schr\"odinger equations but also constitute the orthonormal basis of a Hilbert space which is isomorphic to the space of states of scalar quantum theory
in energy eigenvalue representation
with normalization condition
\beq
 \int \left|\psi\right|^2 dq = 1.
 \label{eq:Constraint4b}
\eeq

Let us call an arbitrary element $\psi \in \H$ of this space a possible external influence. This possible external influence is in general
a linear combination of the basis elements, which we consider as actual external influences by virtue of their role in laboratory space expressed in (\ref{eq:VarProb_1}) and the consistency condition (\ref{eq:HamJac}).
The possible external influences also fulfil the Schr\"odinger equation by linearity of (\ref{eqTDSG}) and from this end we are in formal alignment with
standard quantum mechanics. The difference, however, is that the actual external influence $\psi$  must fulfil the consistency condition (\ref{eq:HamJac}) to be observable in laboratory space, which has an important consequence. 
We illustrate this shortly for the example of a non-degenerate Hilbert space.
Suppose we have a general normed possible external influence given as a superposition of eigenstates and subject to the normalization
(\ref{eq:Constraint4b}).
We assume further that this external influence is observable in laboratory space. Then, this external influence fulfils the consistency
condition (\ref{eq:HamJac}) and its behaviour in laboratory space is given by the equation of motion
of the variational problem (\ref{eq:ConsConClass4}) with constraint (\ref{eq:Constraint4}).
However, the solutions of this problem serve as the basis of the Hilbert space which contradicts our assumption at the beginning
of this derivation. Therefore, a superposition of external influences is not observable in laboratory space, which gives a mathematical explanation of the experimental fact that only energy eigenstates are observable in measurements.
Thus, the measurement problem for non-degenerate Hilbert spaces reduces to a mathematical consequence in our derivation.

Let us highlight  some key points in our derivation which are important for the interpretation of the double-slit experiment.
First, our approach to quantum mechanics emphasizes the energy eigenvalue representation of the Hilbert space. An external influence
is always associated to a specific value of the constant $E$, which is constrained by the experimental setting in laboratory space.
Thus, we can assume that for an external influence in empty laboratory space the energy constant $E$ can have any positive value and our Hilbert 
space $L^2(\rr^3)$ is the space of square integrable function in $\rr^3$.
This fact solves not only the problem of the introduction of
an infinite dimensional Hilbert space in the standard Hilbert space approach to the experiment, see e.{}g.{} \cite{deMuynck1}.
But, as we will see below, gives also a natural framework 
for the positive operator valued measure (POVM) approach, since by Naimark's theorem a POVM is a measure given on a subset
of a standard Hilbert space \cite{deMuynck1}, \cite{Peres1}.

Another key point in our derivation is that the  influence is explicitly external to the physical system which experiences/detects this external influence.
This is reflected in the very beginning of our derivation by the role of the external influence function $g$
in (\ref{eq:VarProb_1})
 as the boundary condition which kicks off the system in laboratory space. The transition to the wave function $\psi$ as the Lie group
 element in (\ref{eq:SubstLog}) is the change of the description about the obtainable information  to the description
 of the object itself.
Thus, in our derivation a quantum object represented by the wave function $\psi$ is explicitly external to the 
physical system in laboratory space represented by the variational problem (\ref{eq:VarProb_1}). 
The Schr\"odinger equation and Hilbert space description  describe the necessary and sufficient conditions in laboratory time and space
to be detectable by the experimental setting represented by (\ref{eq:VarProb_1}).
Therefore, from the perspective of laboratory space our derivation is explicitly non-realistic as the
external influence does not exist in this space until it kicked off the physical system.
Also, our description of dynamics in laboratory space is explicitly non-local which can be seen by the employment
of the Lagrangian (\ref{eq:ConsConClass3}) of the field theory. This Lagrangian uses the unparameterized
space variables $q$ and the "equation of motion" needs to be interpreted in time-unparameterized form. Therefore, the
stationary Schr\"odinger equation (\ref{eq:SG_FinalForm}) is explicitly
non-local in terms of space and time locality as the space coordinates do not depend on the time variable in this formulation. 
Thus, we are in alignment with modern experiments which give strong evidence that quantum theory can not be interpreted
as a local and realist theory but needs to be build as a non-local and non-realist theory \cite{Groeblacher1}, \cite{Legget1}.

However, we need to emphasize that our versions of non-realism and non-locality are more restricted  than the general usage of these terminologies
since this formalism is explicitly local as pointed out in footnote 2 of \cite{Gelfand1}, p68.
Thus, the separation of object space - the space of small external influences - and laboratory space -
the space of the experimental setting - must be regard in the context of local spaces only.
Therefore, in our case, the notions non-realistic and, in particular,  non-local have the paradoxical meaning that they are applied to the local spaces
of a field theory  only,
and reflect our missing knowledge about the structure
of a global manifold similar to the ignorance of the global metric and
global space-time structure in general relativity, which forces us to reject the picture of an absolute space. 
In fact most of the interpretation problems of quantum mechanics are connected to the tacit assumption of an absolute global
space, which we give up in favor to a local description of quantum mechanics.

Moreover, 
the wave function is the representation of the initial disturbing object in our approach and,
therefore, we should regard a wave function as the representation of a quantum event
instead of attaching this function permanently to a quantum system.
This quantum event is closely related to the experimental setting which determines the possible outcomes
perceivable in the laboratory
mathematically expressed in equation (\ref{eq:VarProb_1}).
An advantage of this view is that the objects, e.{}g.{} electrons, can change their quantum and classical roles arbitrarily as the event
of appearance in laboratory space explicitly determines the observability and the visibility of observables, which removes
this interpretation problem from the beginning.

\section{Revisit of the Double-Slit Experiment}

Although the double-slit experiment lies at the heart of quantum mechanics 
it is amazing
how many difficulties one encounters if one tries to actually describe this experiment in terms of the standard
 approach to quantum mechanics. 
 These difficulties can be easily seen as the assumption that the two wave functions $\psi_1$ and $\psi_2$
 originating from the two slits can be regarded as  state vectors of a Hilbert space constituting the basis of
 a two-dimensional Hilbert space only,
 which does not cover the continuity of measured values.
 
Modern approaches use a positive operator valued measure (POVM) to describe the experiment in the language of quantum mechanics, see
e.{}g.{}  (\cite{deMuynck1}, \cite{Peres1}),
which extends the standard measure of quantum mechanics to reflect 
the difference between the number of measurement values and the dimension of the Hilbert space associated to the quantum
system.
But every POVM 
is connected to a standard Hilbert space measure in an extended Hilbert space by Naimark's theorem (\cite{deMuynck1}, \cite{Peres1}),
and thus indicates this approach as the special case of the description  of a quantum effect in a subspace of a larger Hilbert
space. Therefore, the usage of a POVM is an indicator of the lack of knowledge of the correct Hilbert space of
the regarded quantum system.

An alternative approach (\cite{deMuynck1}, p. 343)
assumes that 
the wave functions $\psi_1$ and $\psi_2$ emanating from each slit should be regarded as solutions of the time-independent
Schr\"odinger equation for the same energy $E$ which again leads to the usage of an operator valued measure for the position measurement
in the Hilbert space approach. 
This ansatz is very close to our description of quantum mechanics, which emphasizes the role of the space of external influences with wave
functions subject to the stationary Schr\"odinger  equation (\ref{eq:SG_FinalForm}) for the set of energy eigenvalues $\{E_i\}$.
In general a common fact is that regardless of the used approach the geometrical configuration of the experiment is
crucial to the construction of the Hilbert space and interpretation of the measurement. That is, in every case we deal with two wave functions $\psi_1$ and $\psi_2$ associated to the slits of the experimental setting.

Let us use 
our approach to describe 
the double-slit experiment.
Analogous to the latter approach we regard the wave functions $\psi_{1,j}$ and $\psi_{2,j}$ originating at slit 1 and 2 as
 external influences which kick off a physical system in laboratory space given by (\ref{eq:VarProb_1}) with
energy $E_j$.

The equation of motion (\ref{eq:SG_FinalForm}), i.{}e.{} the stationary Schr\"odinger equation,
of the external influence in laboratory  space is given by the Laplace equations
\beq
\ \ \ \ \ \ \ H \psi_{i,j} = \frac{\partial^2}{\partial q^2} \psi_{i,j}  = E_j \psi_{i,j},
\eeq
which have a continuous spectrum of eigenvalues $E_j$  for each value of the index $i$.
Thus, the vectors  $\psi_{i,j}$  build the basis of an infinite-dimensional Hilbert space avoiding the difficulty of 
a finite eigenvalue spectrum which occurs in the standard approach formulation of the experiment.

Each individual set $\{\psi_{i,j}\}$, $i = 1,2$ constitutes the basis of a Hilbert space $\H_i$ of possible external influences
emanating from slit $i$, which is experimentally realized by opening one slit only during the 
entire  investigation time of the experiment.
The Hilbert space of all external influences for both slits individually measured is then given as the sum of the Hilbert space of the individual slits
\beq
\H = \H_1 \oplus \H_2.
\label{eq:HilberInd}
\eeq

Furthermore, the common set $\S = \{\psi_{i,j}, i = 1,2, H \psi_{i,j} = E_j \psi_{i,j}\}$  of all functions $\psi_{i,j}$ also
constitutes   a basis of a  Hilbert space $\H_{1+2}$  of external influence originating from both slits, which is experimentally realized 
by the usage of a screen as the measuring device. 
This Hilbert space $\H_{1+2}$ is twofold degenerate by virtue of the two wave functions $\psi_{1,j}$ and $\psi_{2,j}$.
It is a well-known fact that every degenerate
Hilbert space with eigenstates $\psi_i$, $i = 1, \ldots, m$ of common eigenvalue $\lambda$ can be represented either in a basis of the individual eigenvectors $\psi_i$  or in a basis given by a superposition of these eigenvectors.
While the individual basis vectors are again equivalent to  external influences originating from an individual slit,
the  basis vectors given as the superposition
\beq
\psi_{j} = \sum_{i = 1}^2 \alpha_i \psi_{i,j}
\label{eq:BasisSuper}
\eeq
represent an external influence with energy $E_j$ given as the superposition of external influences
emanating from both slits. 
Therefore, the role of the wave functions $ \psi_{i,j}$ as  basis vectors in Hilbert space $\H_1 \oplus \H_2$ or $\H_{1+2}$
and the measured values
depends on the experimental set up as one would naturally expect.

Again, recall the crucial point in our derivation  that 
the Schr\"odinger equations describe the appearance of a small external influence in classical laboratory 
space at a definite point in time. That is,
the geometry of the configuration space of the laboratory at this specific point in time determines the quantum event, i.{}e.{} the action of
the influence, and
needs to be taken into account.
In the case of the double-slit experiment the significant  event is located in the detector behind the slits, which 
gives evidence of the occurrence of the quantum particle in the laboratory.
Since we also restricted ourselves to the family of external influences, which are traceable as boundary conditions
in time given by (\ref{eq:Trace1}) and (\ref{eq:Trace2}), 
 the 'ray visibility'  at the point of contact with laboratory space plays a crucial role
in our approach.
Therefore, we have three possibilities of ray visibility which match to the  experimental settings of full visibility,
partial visibility and individual visibility of the slits as indicated in  figure 1, which lead to
the realizations of the Hilbert spaces given above.
\begin{figure}%
\begin{center}
	\includegraphics[width=1.00\textwidth]{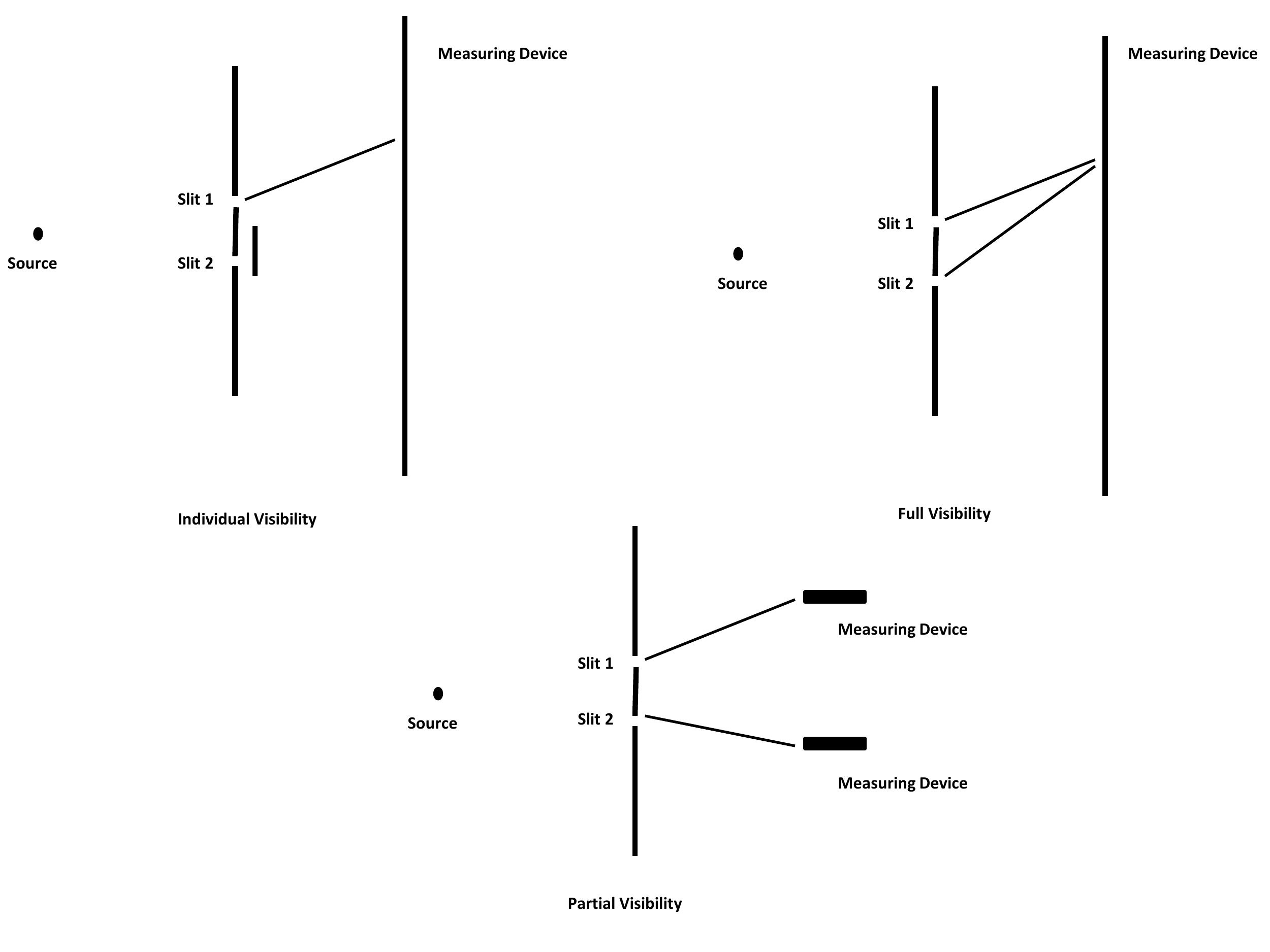}
	\label{Figure1}
	\caption{Possible ray visibilities for different experimental settings}
\end{center}
\end{figure}

Furthermore, the normalization condition (\ref{eq:Constraint4b}) for a state vector in a Hilbert space
also plays  the role
of the constraint (\ref{eq:Constraint4}) for the variational problem (\ref{eq:ConsConClass4}), and we interpret this constraint
as the condition of certain occurrence of the external influence in laboratory space. 
Additionally, only solutions for a specific energy eigenvalue $E_j$  of the variational problem
are observable in an individual measurement
process
by virtue of the consistency condition (\ref{eq:HamJac}),
and every detectable external influence must be an element of the energy eigenvalue basis.
Therefore, energy eigenvalue superpositions 
are not detectable in laboratory space as confirmed by all quantum experiments.

Thus, in the case of the double-slit experiment with one slit closed for each measurement process,
the Hilbert space of possible external influences  $\H_1 \oplus \H_2$ is the sum of the individual Hilbert space of external influences
emanating from one individual slit, and the normalization conditions
\beq
\int \left|\psi_i\right|^2 dq = 1, \ \ \ \ \ i = 1,2
\label{eq:Constraint4b_neu}
\eeq
correspond to the certain event of appearance in laboratory space of the external influence originating from slit $i$ with given energy
$E_j$, where we suppress the energy index $j$ in the following equations.
Thus, the integral (\ref{eq:Constraint4b_neu}) over entire laboratory space expresses the certain appearance 
 of an external influence in laboratory space
originating from the individual slits.
The integral over the interval $[x, x + \delta x]$ gives then the probability 
\beq
p_i(x) = \int_{x}^{x + \delta x} \left|\psi_i(q)\right|^2 dq
\label{eq:PropIntRep}
\eeq
to measure the external influence emanating from
slit $i$ in this region 
and the total probability is the sum of the individual probabilities normed to 1, which is equivalent to the particle picture of standard quantum mechanics.

Let us consider now the Hilbert space of possible external influences originating from both slits.
As mentioned above, this Hilbert space is degenerate and can be represented in two different energy eigenvalue bases.
One possibility is to represent the states in the basis of individual functions $\psi_{1}$ and $\psi_{2}$
with energy eigenvalue $E_j$, which corresponds to the
scenario of partial visibility as shown in the figure above. Then,
each basis element $\psi_{i}$ is again subject to the consistency condition for the observability of a small external influence
in laboratory space with normalization condition (\ref{eq:Constraint4b})
\beq
\int \left|\psi_{i}\right|^2 dq = 1, \ \ \ \ \ i = 1,2.
\eeq
This leads again to the probability (\ref{eq:PropIntRep}) of finding the particle in the interval $[x, x + \delta x]$
with energy $E_j$.
But, analogous to the non-degenerate case a superposition state  $\psi = \alpha_1 \psi_{1} + \alpha_2 \psi_{2}$
is not observable in laboratory space in this basis. 
Therefore, this basis
 describes exclusively the particle picture of the theory.

On the other hand, we can use 
the superposition state 
\beq
\psi = \left( \alpha_1 \psi_{1} + \alpha_2 \psi_{2} \right)
\eeq
as the basis of the energy eigenvalue representation of the Hilbert space, which 
corresponds to full visibility of both slits at the instance in time and space of the quantum event.
This basis allows us to identify
the superposition state of the external influences with
the normalization condition  (\ref{eq:Constraint4b}) which leads to the 
probability
\beq
p_i(x) = \int_{x}^{x + \delta x} \left| \alpha_1 \psi_{1} + \alpha_2 \psi_{2} \right|^2 dq
\label{eq:PropWavRep}
\eeq
 of finding the particle in the interval $[x, x + \delta x]$
with energy $E_j$.
Since the $\psi_{i}$ are eigenstates of the degenerate eigenvalue $E_j$ these are not orthogonal and
result into interference terms leading to the wave interpretation of the experiment.
Now, the individual wave functions $\psi_{i}$ are superposition states  and are not
observable in this basis by virtue of the consistency condition.
Thus, the complementary of wave and particle behaviour is a result of the basis change in a degenerate Hilbert space and
no additional assumptions need to be introduced to explain this experimental fact.

\section{Conclusion}

Our view of quantum mechanics does not only allow the derivation of the Schr\"o-dinger equation and the Hilbert space
of states from physically intuitive assumptions, but gives also an explanation of wave-particle duality. 
 In contrast to standard quantum mechanics we do not need to incorporate additional concepts to explain the  complementarity
 of waves and particles,
 but interpret this fact as the result
of a basis change in a Hilbert space, which is closely tied to a quantum event of an experimental setting in a laboratory.

In addition, all considerations above apply to delayed choice experiments, as experimentally realized in e.{}g.{} \cite{Jacques1}, 
since our approach  singles out for the quantum event of significance
the laboratory space configuration at the definite point in time of appearance in the detector, which removes the paradox that a quantum objects
changes its wave or particle nature during its experimental lifetime. 
The interpretation problems in traditional approaches stem from the implicit assumption of historic consistence
which is connected to the tacit picture that the wave or particle property is attached to an object "flying" from the
source to the detector, which must therefore change its property during the "flight".
In our case the time traceability of the quantum object, the small external influence,
 is reflected by the restriction to the family of consistent
boundary conditions, which gives us the possibility to identify the quantum object at an arbitrary point in time 
by measuring in classical laboratory space. Thus, we can picture the same proposition that if we measured an electron
at time $t_1$ we can conclude it is the same electron we would have measured at point $t_2$ somewhere else in the experimental set up.
The difference, however, is that the small influence, which causes the quantum event, is explicitly external until
its appearance in classical laboratory space
and only the geometry of the classical configuration space at this point in time plays a crucial role.
Therefore, we associate the wave or particle property of an object to a quantum event which is located at a specific
point in time and space in laboratory space, which removes the paradoxical assumption that the particle must change its property
over time.

However, this comes at the cost of a radical modification of quantum mechanics which is more on a conceptual level than a formal one.
First, a quantum event is not an independent property of an object, but is always closely related to the experimental setting in laboratory space since the Hamilton-Jacobi equation for the external influence explicitly requires the Hamiltonian of the experimental environment.
This is not a serious restriction as it enables us to show that the reduction of a state vector 
is equivalent to the transition of a possible external influence to an actual external influence which must be a basis element of the energy
representation of the corresponding Hilbert space.
 In fact it enables us to show  superposition states can only be measured
 for states of a degenerate energy eigenvalue and
 cannot be measured in
an energy superposition state for different energies.

Second, the separation of the external influence from the laboratory system and the view of the wave function as a field in coordinate
space only   incorporate non-reality and non-locality into our derivation.
Moreover, the explicit local construction of our approach and the
separation between object space and laboratory space also removes the implicit assumption of absolute space 
given in the standard approach and introduces the concepts of non-reality and non-locality in a relative context between
local spaces.
We believe that our explicit incorporation of non-locality and non-reality into a local theory can be a first step
to a better physical understanding of quantum mechanics as modern investigations and experiments give evidence that one has to give up
the concepts of reality and locality in quantum mechanics  \cite{Groeblacher1},\cite{Legget1}, \cite{Bell1}, \cite{Aspect1}.

The author wants to thank M. Harris for reading the manuscript.


\end{document}